\documentclass[a4paper,8pt]{article}

\newcommand{\sh}[1]{#1\hskip-5pt /}

\usepackage{latexsym}
\usepackage{a4wide}
\usepackage{graphicx, subfigure} 
\usepackage{cite} 
\usepackage{tabularx}
\usepackage{array}
\usepackage{graphics}     
\usepackage{amsmath}
\usepackage{amssymb}
\usepackage{longtable} 
\usepackage{verbatim} 
\usepackage[table]{xcolor}
\usepackage{booktabs}
\usepackage{hyperref} 
\usepackage[utf8]{inputenc}
\usepackage{multirow}

\definecolor{light-gray}{gray}{0.90}

\begin{document}


\def\thefootnote{\fnsymbol{footnote}}
 
\begin{center}

\vspace{3.cm}

{\Large\bf Resolved Power Corrections to the Inclusive Decay $\bar B \to X_s \ell^+\ell^-$\footnote{Based on talks by T.H. at the Sixth Workshop on Theory,  "Phenomenology and Experiments in Flavour Physics", Capri, 11-13 June 2016, 
and at the MIAPP Scientific Programme "Flavour Physics with High-Luminosity Experiments", Munich, 24 October -18 November 2016.}}

\setlength{\textwidth}{11cm}
                    
\vspace{2.cm}
{\large\bf  
Tobias Hurth$^{,a}$,
Michael Fickinger$^{,a}$,
Sascha Turczyk$^{,a}$,
Michael Benzke$^{,b}$
}
 
\vspace{1.cm}
{\em $^a$PRISMA Cluster of Excellence and  Institute for Physics (THEP)\\
Johannes Gutenberg University, D-55099 Mainz, Germany}\\[0.2cm]
{\em $^b$ I. Institute for Theoretical Physics, University Hamburg\\ 
Luruper Chaussee 149, D-26761 Hamburg, Germany}\\[0.2cm]

\end{center}

\renewcommand{\thefootnote}{\arabic{footnote}}
\setcounter{footnote}{0}

\vspace{1.cm}
\thispagestyle{empty}
\centerline{\bf ABSTRACT}
\vspace{0.5cm}
{
We  identify  the correct power counting of all the variables in the low-$q^2$ window of the inclusive decay $\bar B \rightarrow X_s \ell^+\ell^-$  within the effective theory SCET 
if a hadronic mass  cut is imposed. Furthermore we  analyse the resolved power corrections at the  order $1/m_b$ in a systematic way.  As a special feature,  the resolved  contributions stay nonlocal when the hadronic mass cut is released. Therefore they represent an irreducible uncertainty independent of the hadronic mass cut.
}

\newpage

\section{Introduction}
\label{sec:intro}
As regards  the theoretically clean modes of the indirect search for new physics by means of  flavour observables, 
the inclusive decay mode $\bar  B \to X_s \ell^+\ell^-$ plays a crucial role (for reviews see 
Refs.~\cite{Hurth:2003vb,Hurth:2010tk,Hurth:2012vp}). This inclusive decay mode provides a nontrivial crosscheck of the so-called LHCb anomalies within the recent LHCb  data  on the corresponding exclusive mode~\cite{Aaij:2013qta,Aaij:2015oid}. As demonstrated  in Refs.~\cite{Hurth:2013ssa,Hurth:2014zja}, the future measurements of the inclusive mode will be capable to resolve these puzzles.

A comparison between the inclusive the $\bar B \rightarrow X_s \gamma$ and the inclusive 
$\bar B \rightarrow X_s \ell^+ \ell^-$ decay reveals that the latter is a complementary and a more complex test of the SM, given that  different perturbative electroweak contributions add to the decay rate. As a three body decay process it also offers more observables. Because of  the presence of the lepton-antilepton pair, more structures contribute to the decay rate and some subtleties in the formal theoretical description arise.
This inclusive mode is generally assumed to be  dominated by perturbative contributions like the inclusive $\bar B \to X_s \gamma$ decay if  one eliminates $c \bar c$ resonances with the help of kinematic cuts.
 Research regarding these perturbative contributions has been undertaken extensively and has already reached a highly sophisticated level. The latest analysis of all angular observables in the $\bar B \rightarrow X_s \ell^+\ell^-$ decay has been presented  
in Ref.~\cite{Huber:2015sra}. It  contains  all available perturbative NNLO QCD, NLO QED corrections and also includes the {\it known}  subleading power corrections.

For the inclusive modes $\bar B \rightarrow X_s \gamma$ and $\bar B
\rightarrow X_s \ell^+ \ell^-$, it is possible to demonstrate  that,  {if} only the leading operator in the effective
Hamiltonian (${\cal O}_7$ for $\bar B \to X_s \gamma$, ${\cal O}_9$ for $\bar B \to X_s \ell^+\ell^-$) 
is taken into account, the heavy mass expansion (HME)  allows for the calculation of 
the inclusive decay rates of a hadron containing a heavy quark,
especially a $b$ quark~\cite{Chay:1990da,Bigi:1992su}.  
In this case,  one arrives at a local operator product expansion (OPE) based on the optical theorem. The free quark model represents  the first term in the constructed expansion in powers of
$1/m_b$ and is  therefore the dominant contribution.  In the applications to
inclusive rare $B$ decays, one finds no correction of order $\Lambda/m_b$
to the free quark model approximation within this OPE because of the equations of motion. As a consequence, the corrections to the
partonic decay rate begin  with $1/m_b^2$ only. This implies
that there is a small numerical impact of the nonperturbative corrections on 
the decay rate of inclusive modes.

However,  there are more subtleties to be taken into account if other than  the leading operators are considered.
As already demonstrated in Ref.~\cite{Ligeti:1997tc}, there is no OPE for the
inclusive decay $\bar B \rightarrow X_s \gamma$ if one analyses 
operators beyond the leading electromagnetic dipole operator ${\cal O}_7$.
Indeed, there are  the so-called resolved photon contributions. These  include  subprocesses in which the photon connects  to light partons
instead of coupling   directly to the effective weak-interaction vertex~\cite{Voloshin:1996gw,Lee:2006wn}.
Within the inclusive decay   $\bar B \to X_s \gamma$,     a systematic analysis~\cite{Benzke:2010js} of
all resolved photon contributions related to other operators in the weak
Hamiltonian establishes this breakdown of the local OPE within the
hadronic power corrections as a generic result.  Within soft-collinear effective  theory (SCET),
an analysis of such linear power corrections is possible.
Clearly, one has to confront difficulties if estimating such nonlocal matrix elements. An irreducible theoretical uncertainty of $\pm
(4-5)\%$ for the total $CP$ averaged decay rate, defined with a
photon-energy cut of $E_\gamma = 1.6$ GeV, cannot be eliminated~\cite{Benzke:2010js}.

In the present letter the resolved contributions to  the inclusive decay $\bar B \rightarrow X_s \ell^+ \ell^-$ are studied. 
Within the inclusive decay {\bf $\bar B \to X_s \ell^+ \ell^-$}, 
the hadronic ($M_X$) and dilepton invariant ($q^2$) masses are independent 
kinematical quantities.  An invariant mass cut on the hadronic final state system ($M_X \lesssim 2\,\text{GeV}$) is necessary in order to suppress potential huge backgrounds. This cut implies  no additional constraints in the high-dilepton-mass region. In the low-dilepton region, the cut on the hadronic mass leads to  a specific  kinematics in which the standard OPE  collapses and  one has to introduce nonperturbative $b$-quark distributions, so-called shape functions. Given the specific kinematics of low dilepton masses $q^2$ and of small hadronic masses $M_X$, one has to deal with  
a multi-scale problem for which soft-collinear effecive theory (SCET) is the appropriate tool.

A former SCET analysis made use of the universality of the leading shape
function to show that the reduction resulting from  the $M_X$-cut  can be calculated  for  all angular observables of  the inclusive decay $\bar B\rightarrow X_s \ell^+ \ell$. The  effects of subleading shape functions
imply  an additional uncertainty of $5\%$~\cite{Lee:2005pk,Lee:2005pw}. A later analysis~\cite{Lee:2008xc}
estimates the uncertainties due to subleading shape functions more
conservatively. In the future it may be possible to
decrease such uncertainties significantly by constraining both the
leading and subleading shape functions using the combined $B \to
X_s\gamma$, $B \to X_u\ell \bar\nu$ and $B \to X_s \ell^+\ell^-$
data~\cite{Lee:2008xc}.

All these former analyses, however,  are based on the problematic assumption, that  $q^2$ is  a hard scale in the kinematical region of low $q^2$ and of small $M_X$.
 By contrast,  our present  SCET analysis will demonstrate explicitly that the hadronic cut  implies  the scaling of $q^2$ being  not hard but (anti-) hard-collinear in the low-$q^2$ region.

Therefore it is the first and primary task to identify  the correct power counting of all the variables in the low-$q^2$ window of the inclusive decay $\bar B \rightarrow X_s \ell^+\ell^-$  within the effective theory SCET  in case a 
hadronic mass  cut is imposed. Moreover, the resolved power corrections have to be examined systematically.   As already mentioned, in these contributions the photon couples to light partons instead of connecting directly to the effective weak-interaction vertex.  There would be no such contribution, if  $q^2$ was hard in SCET. 
 Furthermore, we will show that the resolved  contributions have the special feature  that  they stay nonlocal when the hadronic mass cut is released. In this sense they thus ead to  an irreducible uncertainty that is not dependent on the hadronic mass cut.

\section{SCET analysis of the resolved contributions}
The effective operator basis for the underlying parton interaction of the semileptonic flavour changing neutral current decay $\bar B \to X_s \ell^+\ell^-$ is well-known~\cite{Buchalla:1995vs}.  Many higher-order calculations have led to the availability of NNLO precision and NNLL resummation in the strong coupling $\alpha_s$. At the relevant scale $m_b$ of the $b$-quark, all heavier fields are integrated out. The effective operator basis only includes only active flavours.

When calculating the inclusive decay mode $\bar B \to X_s \ell^+\ell^-$, one is confronted with two problems:  The first problem stems from the fact that the integrated branching fraction is dominated by
resonant $q\bar q$ background, in particular with $q=c$, i.e. resonant $J/\psi
\rightarrow \ell^+ \ell^-$ intermediate states for the (virtual) photon, which exceeds the nonresonant
charm-loop contribution by two orders of magnitude. This phenomenon should not be misinterpreted as a striking failure of global parton-hadron duality as shown in Ref.~\cite{Beneke:2009az}.
In any case, $c \bar c$ resonances appearing  as large peaks in the dilepton invariant mass spectrum are eliminated  by appropriate  kinematic cuts -- leading to so-called `perturbative $q^2$-windows', namely  the low-dilepton-mass  region $1\,{\rm GeV}^2 < q^2 = m_{\ell\ell}^2   < 6\,{\rm GeV}^2$, and also the high-dilepton-mass region with $q^2 > 14.4\,{\rm GeV}^2$.  

The second problem to be faced  is related to the fact that in  a realistic experimental environment one has to suppress potential huge backgrounds by an invariant mass cut on the hadronic final state system ($M_X \lesssim 2\,\text{GeV}$). This cut does not involve any  additional constraints in the high-dilepton-mass region. But in the
low-dilepton mass region we have in the $B$ meson rest frame due to $q= p_B -p_X$:
\begin{equation} 
2\, M_B\, E_X \, = M_B^2 +M_X^2 -q^2\,.
\end{equation}
Thus, for low enough $q^2$ in combination with $M_X^2 \ll E_X^2$,  the $X_s$ system is jet-like with  $E_X  \sim  M_B$. This also means that  $p_ X$ is near the light cone.

Considering these kinematic constraints, soft-collinear-effective theory (SCET)~\cite{Bauer:2001yt} is the adequate tool to analyse  the factorization properties of inclusive $B$-meson decays in this region and to 
examine  the multi-scale problem. Thus, the cuts in the two independent kinematic variables, namely the hadronic and dilepton invariant masses, force us  to study the process in the so-called shape function region
with a large energy $E_X$ of order   $M_B$ and low invariant mass  $ M_X \sim \sqrt{m_b \Lambda_\text{QCD}}$ of the hadronic system. SCET makes it possible to get systematically hold of  a scaling law of the momentum components. In this set-up the scales $\Lambda_\text{QCD}$, $M_X$, $q^2$ and $M_B$ are relevant. 
One arrives at the following hierarchy for the ratio of the scales:
\begin{equation}
  \Lambda_\text{QCD}   /  {M_B}    \ll   M_X  /   M_B      \ll 1\,.
\end{equation}  
Therefore it is of high relevance to resum logarithms between these scales.  One can resum  the logarithms of these scale ratios systematically. What is more, one can  factorize the effects resulting from diverse regions. In this way, one can calculate the process in a consistent expansion and factorize off effects that can be calculated
perturbatively. This reduces the non-perturbative quantities to a limited set of soft functions. If one defines $\lambda = \Lambda_\text{QCD}/M_B$, one numerically finds that  $M_X \lesssim \sqrt{M_B \Lambda_\text{QCD}} \sim M_B \sqrt{\lambda}$. This implies the power-counting scale for the possible momentum components in 
light-cone coordinates $n^\mu = (1,0,0,1)$ and $\bar n^\mu = (1,0,0,-1)$.  Any four-vector may be decomposed  according to  $a^\mu = n\cdot a \,\,\bar n^\mu/2 + \bar n\cdot a \,\, n^\mu/2 + a_\perp^\mu\,.$
The short-hand notation is defined to be  $a \sim (n\cdot a, \bar n\cdot a, a_\perp)$ in order to specify the scaling of the momentum components in powers of $\lambda$. Within the validity of SCET,  we have a hard momentum region $p_\text{hard} \sim (1, 1, 1)$, a hard-collinear region $p_\text{hc} \sim ( \lambda, 1 , \sqrt{\lambda})$, an anti-hard-collinear region $p_{\overline{\text{hc}}} \sim ( 1, \lambda , \sqrt{\lambda})$, and a soft region $p_\text{soft} \sim (\lambda, \lambda, \lambda)$. 
\begin{figure*}[t!]
\begin{center}
\includegraphics[scale=0.37]{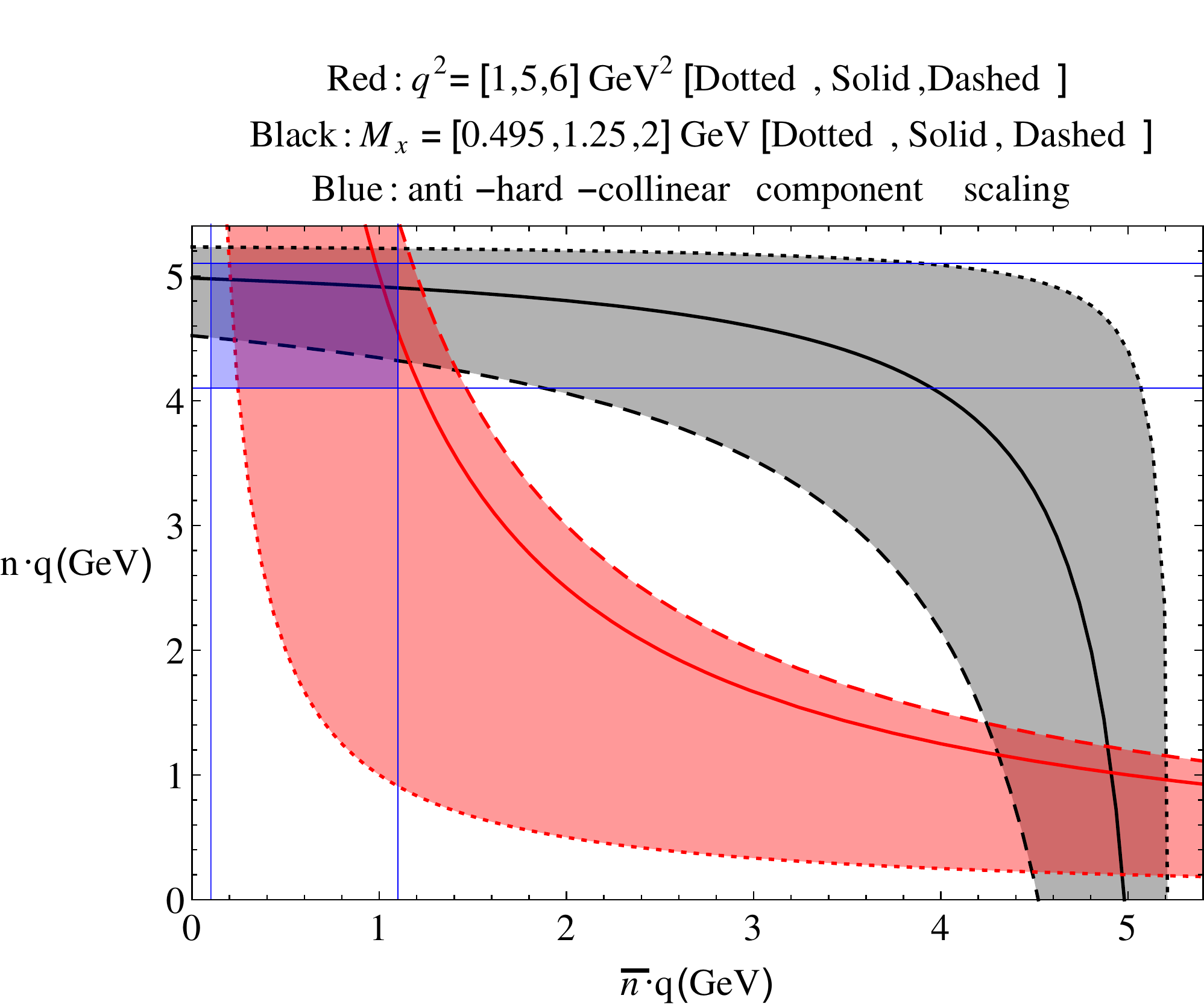}\hspace{0.5cm}\includegraphics[scale=0.37]{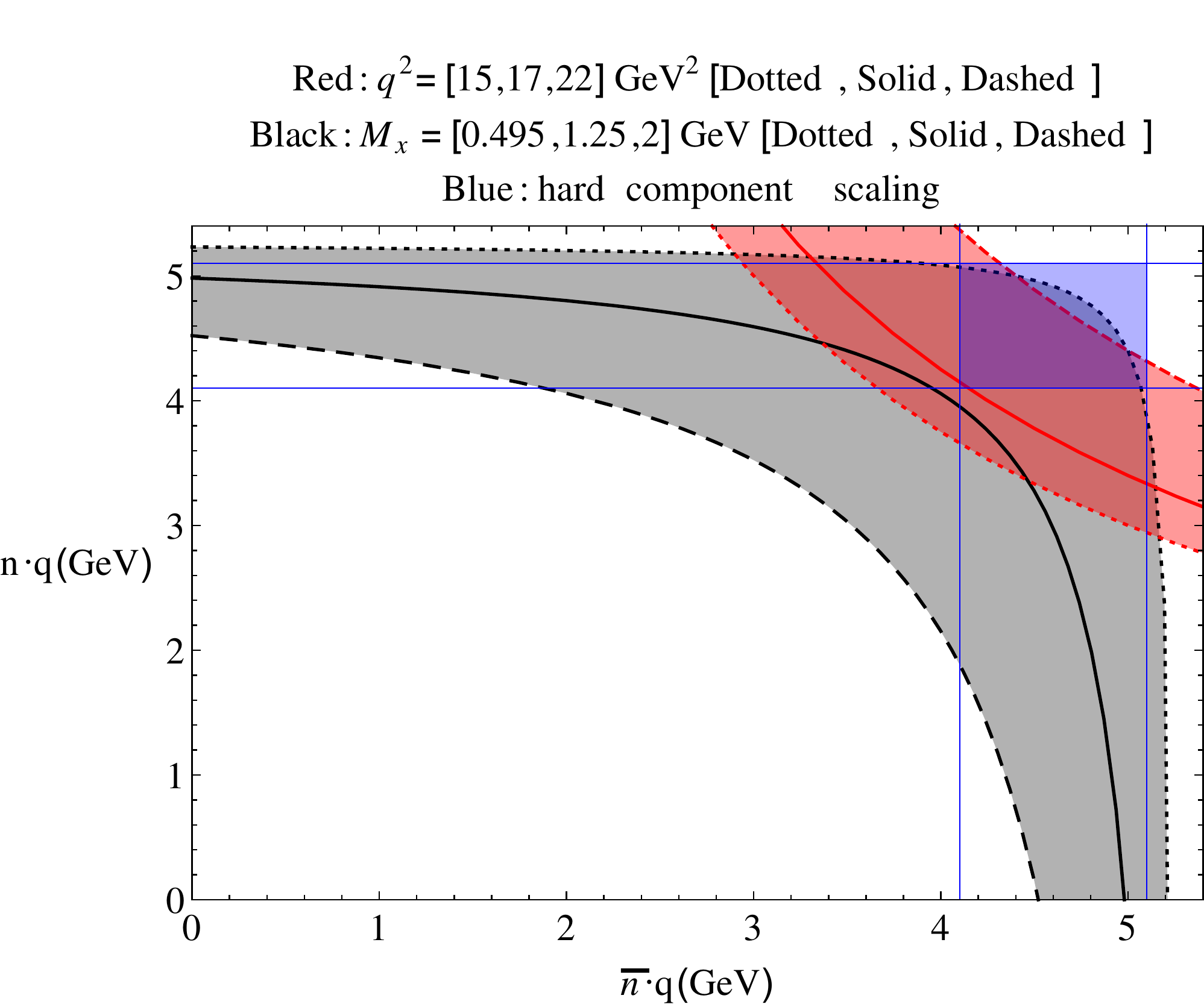}
    \caption{$q^2 = (n \cdot q) (\bar n \cdot q)$ with $q_\perp = 0$ for the two perturbative mass windows. The gray band shows the experimental hadronic invariant mass cut with the $K$ as the lowest mass state. The red band corresponds to the $q^2$ cut. The blue lines indicate the scaling of the two light-cone components. Left: Low invariant mass window. The scaling of $q_{\overline{\text{hc}}}$ is indicated. Right: High invariant mass window, with the maximally allowed value of $M_B$. The scaling of $q_\text{hard}$ is indicated.}\label{Fig:scaling_law}
\end{center}
\end{figure*}

Regarding  the two-body radiative decay, the kinematics requires  $q^2 = 0$ and $E_\gamma \sim m_b/2$.  The scaling including the invariant mass and photon energy requirement is fixed to be a hard-collinear hadronic jet recoiling against an anti-hard collinear photon. 

In the case of a lepton-antilepton pair in the final state, one has to pose a restriction on the momentum transfer to the leptons around the mass window of the $c\bar c$ resonances as described above. In Figure~\ref{Fig:scaling_law} a comparison is made between  the momentum scaling of the lepton-antilepton pair in terms of the light-cone coordinate decomposition and the experimental cuts. The gray band corresponds to the hadronic invariant mass cut in order to suppress background, while the red band is the $q^2$ constraint to reject the $c\bar c$ resonances. The blue lines show the validity of SCET in terms of the momentum component scaling, on the left figure for an anti-hard-collinear scaling, while on the right one for a hard momentum scaling. We note that there are two  solutions for the left figure, as we may view the leptons to be anti-hard-collinear and the hadronic jet collinear and vice versa.
The high mass window corresponds to hard leptons. It is beyond  of the validity of a description in terms of SCET. It is
obvious that the current mass cuts does not influence this scenario --  in contrast to the low $q^2$ region. The overlap of the red and gray band corresponds to the allowed region after experimental cuts. It is in good agreement with our assumptions for the effective theory  which is approximately given by the blue rectangle. Thus, when assigning an anti-hard-collinear momentum to the virtual photon and a hard-collinear one to the hadronic system, one arrives at   a good approximation in the validity window of both the experimental requirement and the effective theory. 

\begin{figure*}[t!]
\begin{center}
    \centering\includegraphics[scale=0.37]{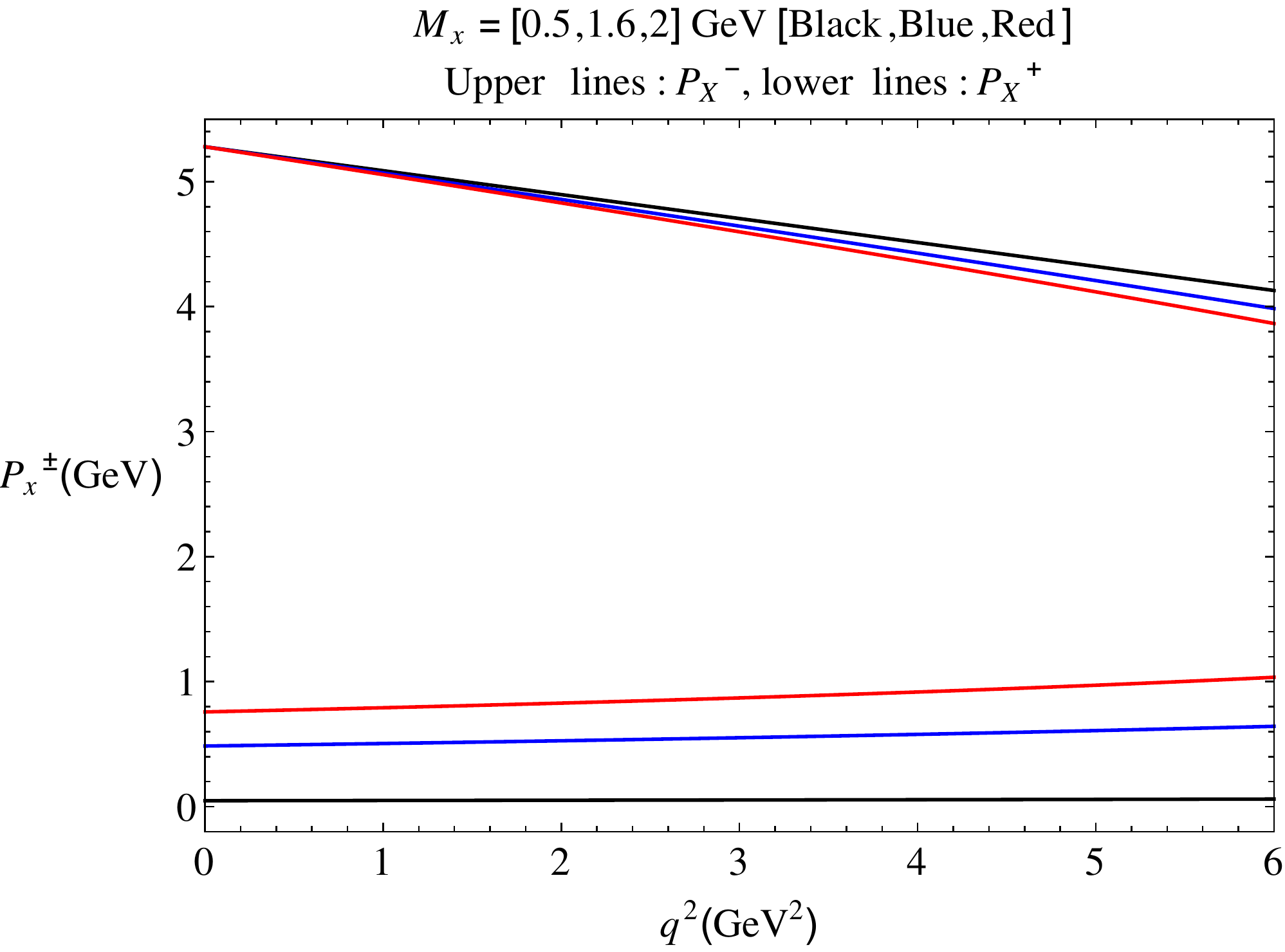}\hspace{0.5cm} \includegraphics[scale=0.37]{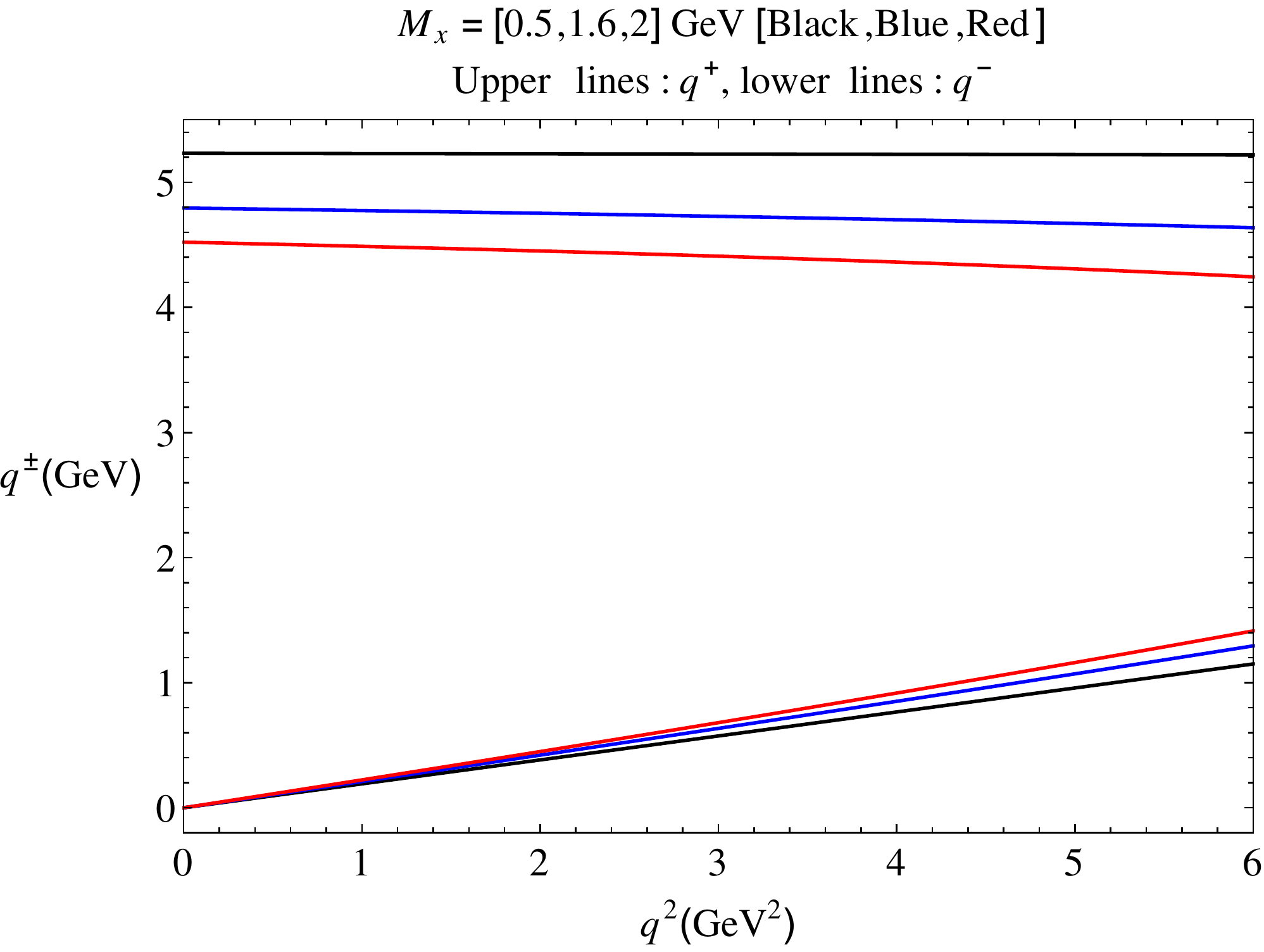}
    \caption{The scaling of the momentum components of the hadronic system $p_X^+ = n\cdot p_X$ and $p_X^- = \bar n \cdot p_X$ [left] and the lepton system $q^+ = n\cdot q$ and $q^- = \bar n \cdot q$ is plotted as a function of $q^2$ for each three values of the hadronic invariant mass.}\label{Fig:scaling_law_momentum}
\end{center}
\end{figure*}

In order to demonstrate this in more explicit terms, one can introduce the two light-cone components of the hadronic momentum  with  $p_X^- p_X^+ = m^2_X$ and $p_X^\perp = 0$ via 
\begin{align}
\bar n\cdot p_X  &=    p_X^- =   E_X  +  |  \vec{p}_X |  \sim   {O}(M_B)\nonumber\\  n\cdot p_X  &=    p_X^+ =   E_X  -  | \vec{p}_X |  \sim   {O}(\Lambda_{\rm QCD})\,.
\end{align}
When making use of the kinematical relations, the leptonic light-cone variables are set  by 
\begin{align}
q^+ &= n\cdot   q = M_B - p^+_X\nonumber\\
 q^- &= \bar n\cdot  q = M_B - p^-_X = q^2 / (M_B - p_X^+)\,.
 \end{align}
In  Figure~\ref{Fig:scaling_law_momentum}, we demonstrate that  the scaling  of the momentum components of the hadronic system $p_X^+ = n\cdot p_X$ and $p_X^- = \bar n \cdot p_X$  (left plot) 
and of the lepton system  $q^+ = n\cdot q$ and $q^- = \bar n \cdot q$ (right plot)   as function of $q^2$ for three different values of the hadronic  mass cut.  
Here we rely on the assumption for the experimentally envoked cuts that the effective two-body decay system $B\rightarrow X_s \gamma^*$ is  aligned along the light-cone axis without a perp component. In this case,  the hadronic system scales as hard-collinear, while the lepton system scales as anti-hard collinear.
But one can also extract from the plots that a lower cut of $q^2 \lesssim 5 \text{ GeV}^2$ instead of $q^2 \lesssim 6 \text{ GeV}^2$ is preferred  because a higher value of the $q^2$ cut pushes the small component to values slightly beyond our assumptions of the momentum component scaling and therefore neglected higher order terms may have a more sizable contribution. 
Nevertheless, the  assumption of a hard $q$ momentum as used 
in the calculations of Refs.~\cite{Lee:2005pk,Lee:2005pw,Lee:2008xc} is not appropriate. It also implies both a different scaling and also a different matching of the operators. 

Therefore, we use SCET to  describe the hadronic effects with SCET in correspondence to an expansion of the forward scattering amplitude in non-local operator matrix elements. 
A  factorization formula can be derived. It is completely analogous  to the radiative decay in \cite{Benzke:2010js}:
\begin{align}\label{fact2}
   &d\Gamma(\bar B\to X_s\ell^+ \ell^-)
   = \sum_{n=0}^\infty\,\frac{1}{m_b^n}\, \sum_i\,H_i^{(n)} J_i^{(n)}\otimes S_i^{(n)} +\nonumber \\
   &+ \sum_{n=1}^\infty\,\frac{1}{m_b^n}\,\bigg[ \sum_i\,H_i^{(n)} J_i^{(n)}\otimes S_i^{(n)}\otimes\bar J_i^{(n)} +\nonumber\\
    &\quad\quad \quad\,\,\,\,\,+ \sum_i\,H_i^{(n)} J_i^{(n)}\otimes S_i^{(n)} \otimes\bar J_i^{(n)}\otimes\bar J_i^{(n)} \bigg] \,. 
\end{align}
The formula includes  the so-called direct contributions in the first line, whereas the second and third line describe the resolved contributions which occur first only at the order $1/m_b$ in the heavy-quark expansion.  
$H_i^{(n)}$ are the hard functions describing physics at the high scale $m_b$. $J_i^{(n)}$ are so-called jet functions characterizing the physics of the hadronic final state $X_s$ with the invariant mass in the range described above.
The hadronic physics associated with  the scale $\Lambda_\text{QCD}$ is parameterized by the soft functions $S_i^{(n)}$. Similar to the radiative decay investigated in Ref.~\cite{Benzke:2010js}, we have in addition resolved virtual-photon  contributions in the second line, whose  effects are
described by new jet functions $\bar J_i^{(n)}$. This occurs due to the coupling of virtual photons with energies of order $\sqrt{m_b \Lambda_\text{QCD}}$ to light partons instead of the weak vertex directly. Consequently, they probe the hadronic substructure at this scale.
Resolved effects  may occur as a single or double ``resolved'' contribution due to interference of the various operators, which also have the ``direct virtual-photon'' contribution.
Finally, the soft or shape functions are defined in terms of forward matrix elements of non-local heavy-quark effective theory (HQET) operators. This limited set of shape functions cannot be calculated perturbatively. Nevertheless it leads to 
a  systematic analysis of hadronic effects in this decay mode.
We denote the convolution of the soft and jet function due to the occurence of common variables with the symbol             $\otimes$.  Finally, we note  that this factorization formula cannot be completely proven.  This has already been discussed in Ref.\cite{Benzke:2010js}:  There is one particular case in which a  UV divergent convolution integral exists within the resolved contribution.       
The contribution from ${\cal O}_8 - {\cal O}_8$ includes  a UV divergence canceling  the $\mu$-dependence of the corresponding subleading jet function --  a cancelation that is expected and required. But in order to arrive at a consistent description, one has to use a proper factorization of the anti-jet functions. The convolution of the two anti-jet functions with the soft-function solves the item. To arrive at the correct factorization result, the limit of the DimReg parameter $\epsilon$ must be taken after the convolution has been carried out. However, this contradicts the assumptions in the factorization formula. We note that there are also divergent convolution integrals in SCET in power-suppressed contributions to hadronic $B$ meson decays. The important difference to our present case is that these divergences have an IR-origin.

It is necessary to combine $\text{QCD}\otimes\text{QED}$ in terms of SCET in order to describe the process mentioned. Kinematically, one has to consider the fact that the hadronic part has to be described in terms of SCET for a proper and consistent description. This is also true for QED. We have to describe  the QED fields in terms of an SCET-like theory.
Thus, we examine  the matching of ${\cal O}_7$ onto SCET fields, where we consider the (virtual) photon to be power-counted as well. Then the electromagnetic dipole operator can  then be written as
 \begin{equation}
   {\cal O}_{7}= -\frac{e}{8\pi^2}\,m_b\,
    \bar s\sigma_{\mu\nu}(1+\gamma_5) F^{\mu\nu} b 
\end{equation}
${\cal O}_{7}$ is matched onto the operators following the notation of \cite{Beneke:2002ph} to the leading operator with $\mathcal{A}$ being the Wilson line dressed gauge-invariant photon field. Here we suppress a factor
$-\frac{e m_b}{4\pi^2}\,e^{-im_b\,v\cdot x}$.

\begin{equation}\label{eq:Q7_A}
{\cal O}_{7 A}^{(0)} = \bar{\xi}_{{\rm hc}}\,\frac{\bar{\sh{n}}}{2}\, [i n\cdot \partial \sh{\mathcal{A}}^{{\rm em}}_{\perp}]\,(1+\gamma_5)  h  \,. 
\end{equation}
The scaling pf the photon field is given by $ (n\cdot \mathcal{A}^\text{em}, \bar n\cdot \mathcal{A}^\text{em}, \mathcal{A}^\text{em}_\perp) \sim ( 0, \lambda , \sqrt{\lambda})$. Here gauge invariance implies  $n\cdot \mathcal{A}^\text{em} = 0$ 
even though it is off-shell. The scaling of ${\cal O}_7$ is $\lambda^\frac{5}{2}$. As regards  the semi-leptonic operators, 
\begin{equation}
 {\cal O}_9 = \frac{\alpha}{2\pi} (\bar sb)_{V-A} (\bar \ell\ell)_{V} \, , \, \, \, {\cal O}_{10} = \frac{\alpha}{2\pi} (\bar sb)_{V-A} (\bar \ell\ell)_{A}
 \end{equation}
the matching results into the following SCET operators:
\begin{align}
     {\cal O}_9^{(1)} = \frac{\alpha}{2\pi} ( \bar{\xi}_\text{hc}^s  [1+\gamma^5 ] h) ( \bar{\xi}_\text{hc}^\ell \frac{\sh{n}}{2}  \xi_\text{hc}^\ell), \hspace{0.3cm}
    {\cal O}_{10}^{(1)} =\frac{\alpha}{2\pi} ( \bar{\xi}_\text{hc}^s [1+\gamma^5 ] h) ( \bar{\xi}_\text{hc}^\ell \frac{\sh{n}}{2}  [\gamma^5] \xi_\text{hc}^\ell)\,.
\end{align}
The two operators scale as $\lambda^{\frac12 +\frac32 +2 \frac12} = \lambda^3$, thus, they are  suppressed by $\lambda^\frac12$ against the contribution from ${\cal O}_7$. This feature changes in the high $q^2$ region as in this case the leptons are hard and do not add a power suppression. Thus, the leading order reference is given by ${\cal O}_7 - {\cal O}_7$ at the order of $\lambda^5$. We consider all contributions up to order $1/m_b$ corrections, i.e. terms up to $\lambda^6$. This implies that  we have to take into account only the leading part of ${\cal O}_{9,10} - {\cal O}_{9,10}$ and also  the subleading part of ${\cal O}_7 - {\cal O}_7$. This includes  subleading soft and  jet functions.

\begin{figure*}[t!]
\begin{center}
\includegraphics[scale=0.25]{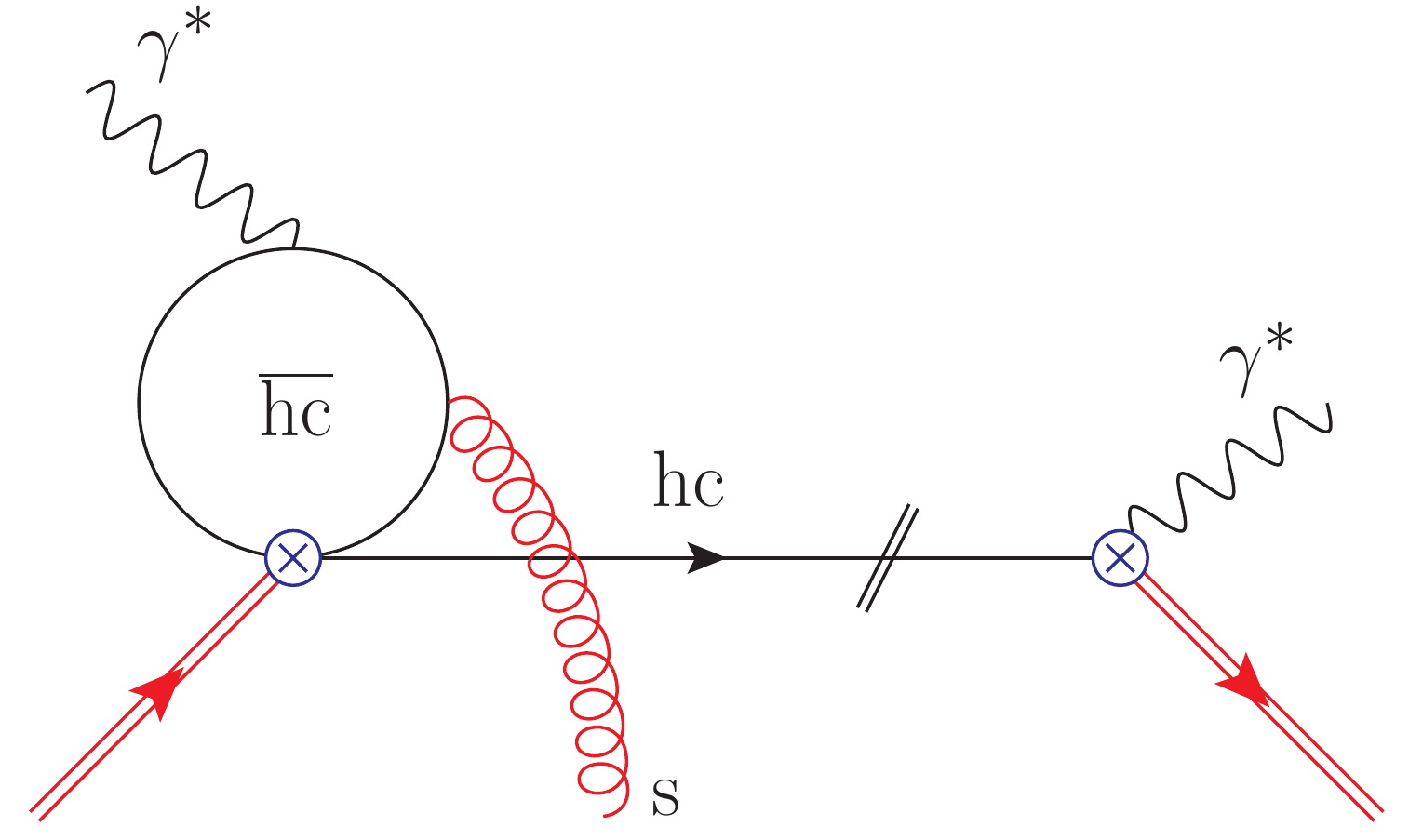}\includegraphics[scale=0.25]{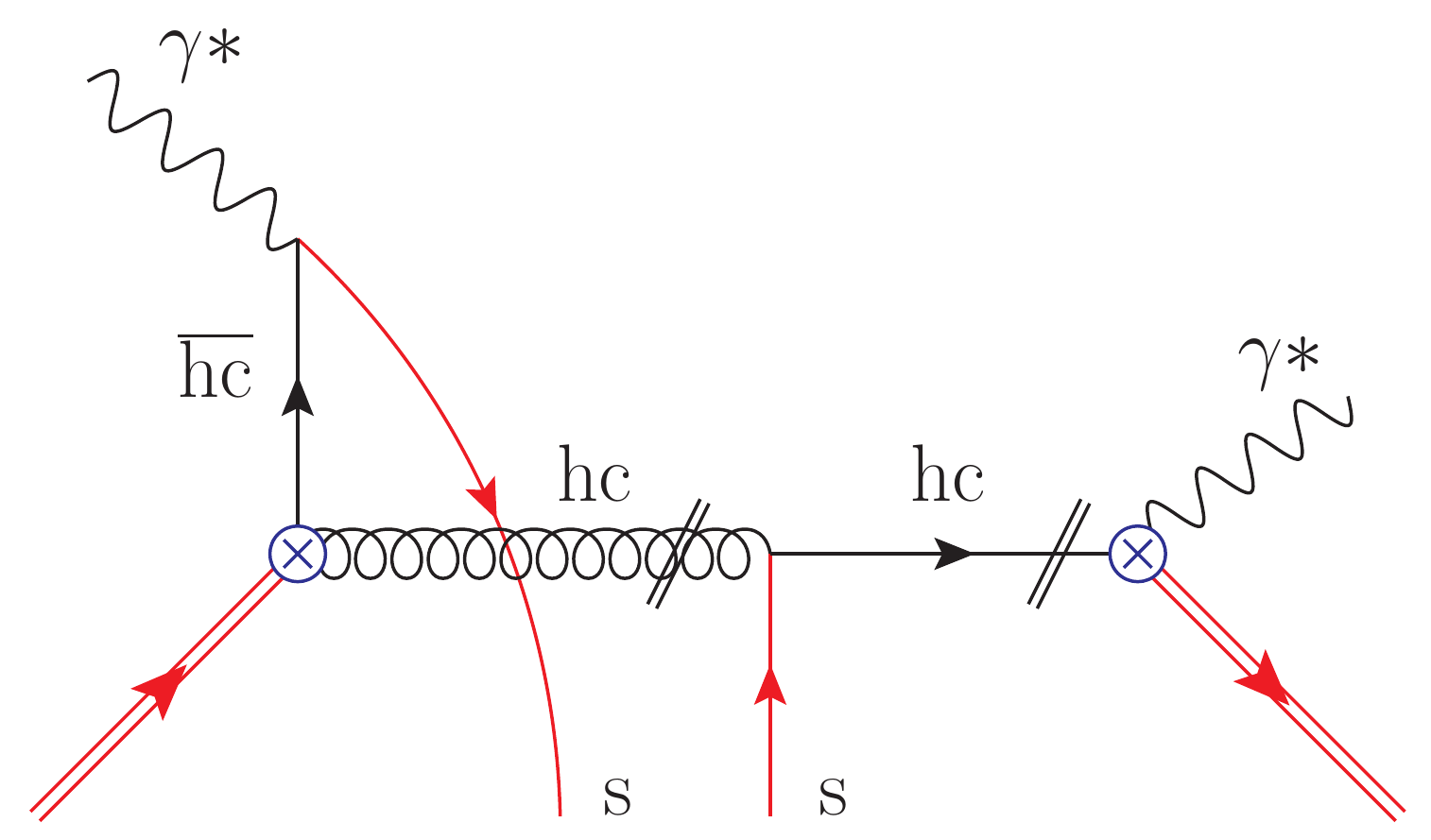}\includegraphics[scale=0.25]{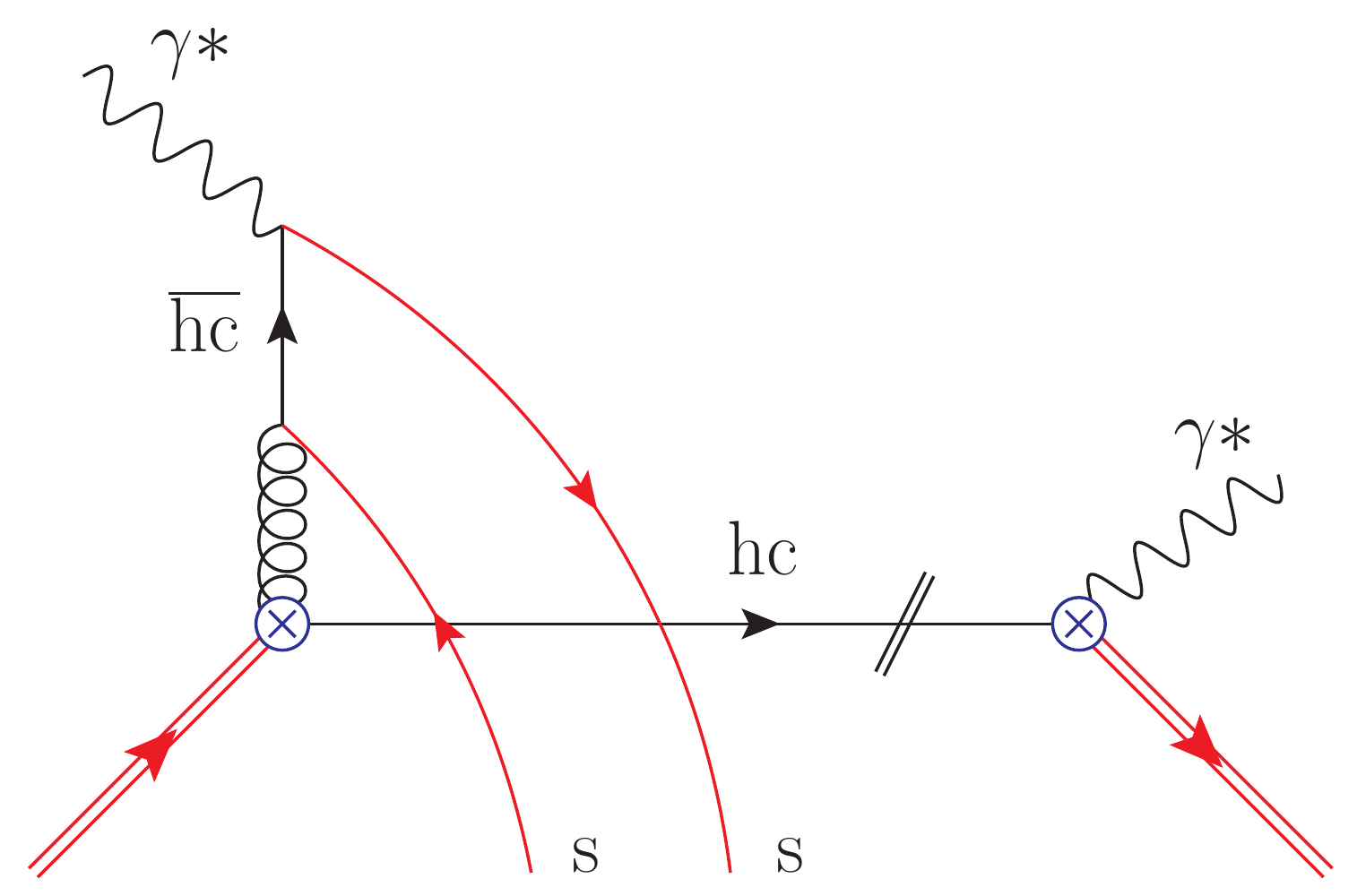}\includegraphics[scale=0.25]{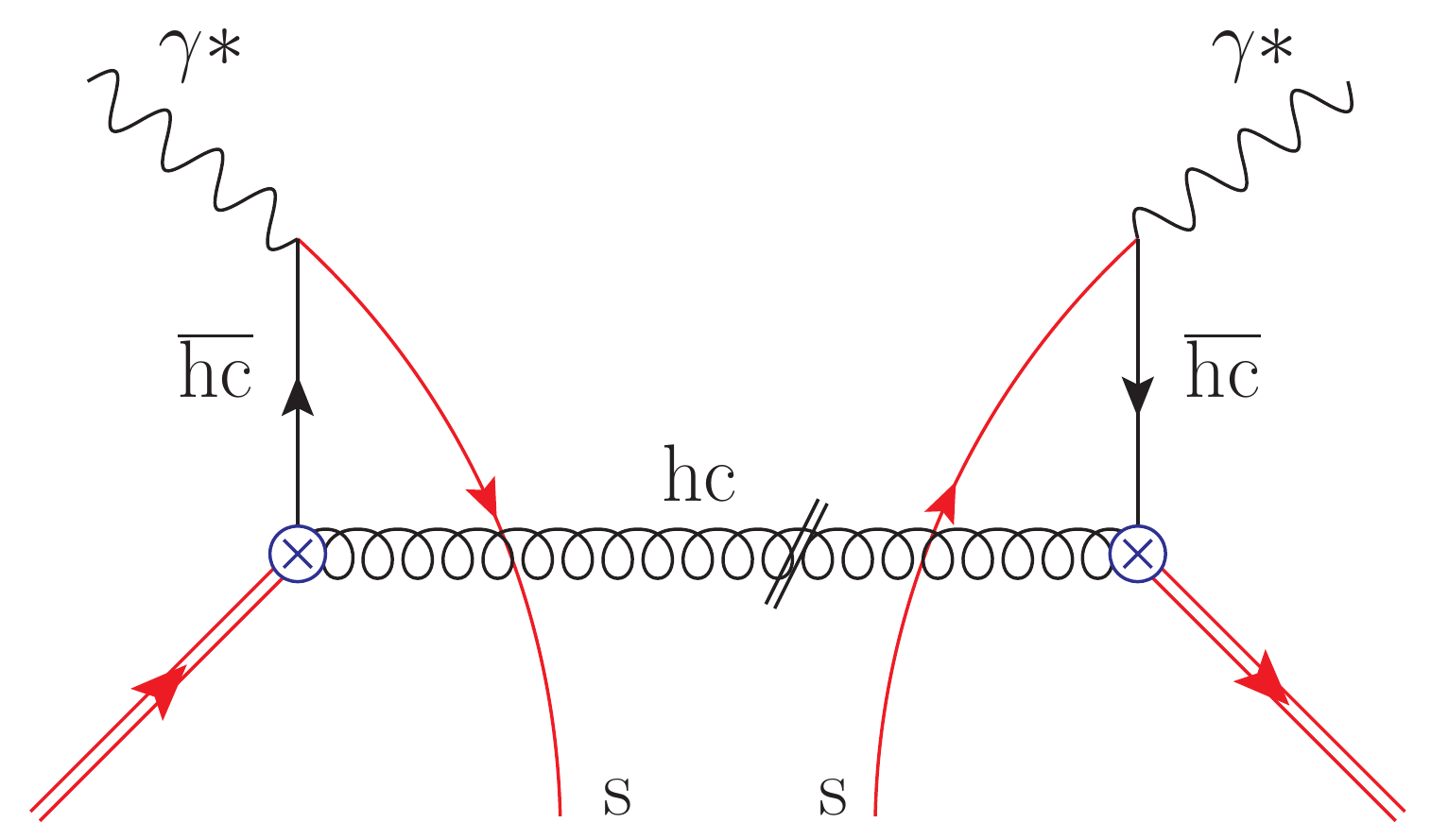}
    \caption{Diagrams arising from the matching of the ${\cal O}_1^q - {\cal O}_{7}$ contribution onto SCET,  of the two ${\cal O}_{7} - {\cal O}_{8}$ contributions, and of the  ${\cal O}_{8}-{\cal O}_{8}$ contribution  (from left to right). Red indicates soft fields, black (anti-) hardcollinear fields. Hard fields have already been integrated out. }  \label{Fig:diagrams}
\end{center}
\end{figure*}

We consider the resolved contributions  to order $1/m_b$. This includes the computation of  the resolved contributions from ${\cal O}_1 - {\cal O}_7$, ${\cal O}_7 - {\cal O}_8$ and ${\cal O}_8 - {\cal O}_8$.  We emphasize the fact  that the conversion of a photon to the lepton pair does not lead to  a further power suppression. 
Figure~\ref{Fig:diagrams} illustrates the various resolved contributions. 

Let us discuss the structure of the ${\cal O}_{8} - {\cal O}_{8}$ contribution in an exemplary mode.  For the differential decay rate one finds: 
\begin{eqnarray}
\frac{d\Gamma}{dn\cdot q\,d\bar{n}\cdot q}\sim \frac{e_s^2 \alpha_s}{m_b}\int d\omega\, { \delta(\omega+m_b-n \cdot q}) 
\int \frac{d\omega_1}{{ \omega_1+\bar n\cdot q+i\varepsilon}}\int \frac{d\omega_2}{{\omega_2+\bar n\cdot q-i\varepsilon}} {g_{88}(\omega,\omega_1,\omega_2)}, 
\end{eqnarray} 
while the shape function has the following structure:
\begin{eqnarray} 
 g_{88}(\omega,\omega_1,\omega_2) 
   = \int\frac{dr}{2\pi}\,e^{-i\omega_1 r}
    \int\frac{du}{2\pi}\,e^{i\omega_2 u}
    \int\frac{dt}{2\pi}\,e^{-i\omega t}  \frac{1}{M_B}  \langle\bar B |\bar h({tn})\dots s({tn+u\bar n})\bar s({r\bar n})\dots h({0})|\bar B\rangle \,.
\label{O8O8}
\end{eqnarray}
There are two remarks in order: First, all diagrams in Figure~\ref{Fig:diagrams} reveal  that if the lepton momenta were assumed to be hard, there would be no resolved contributions: The hard momentum of the leptons would also imply 
a hard momentum of the intermediate parton. The latter would be integrated out at the hard scale and the virtual photon would be connected directly to the effective electroweak interaction vertex. 
Secondly, as demonstrated in Eq.~(\ref{O8O8}),  the shape function is nonlocal in both light cone directions. Therefore, the resolved contributions are still  nonlocal even when the hardronic mass cut is relaxed. In this sense the resolved contributions represent an irreducible uncertainty within the  inclusive decay $\bar B \to X_s \ell^+\ell^-$. 

A complete analysis of all resolved contributions to order $O(1/m_b)$ and their phenomenological impact within the inclusive decay  $\bar B \to X_s \ell^+\ell^-$ can be found  in  Ref.~\cite{Benzke:2017woq}.


\begin{thebibliography}{00}





\bibitem{Hurth:2003vb}
  T.~Hurth,
  ``Present status of inclusive rare B decays,''
  Rev.\ Mod.\ Phys.\  {\bf 75} (2003) 1159
  [arXiv:hep-ph/0212304].

\bibitem{Hurth:2010tk}
  T.~Hurth and M.~Nakao,
  ``Radiative and Electroweak Penguin Decays of B Mesons,''
  Ann.\ Rev.\ Nucl.\ Part.\ Sci.\  {\bf 60} (2010) 645
  [arXiv:1005.1224 [hep-ph]].

\bibitem{Hurth:2012vp}
  T.~Hurth and F.~Mahmoudi,
  ``New physics search with flavour in the LHC era,''
  Rev.\ Mod.\ Phys.\  {\bf 85} (2013) 795
  [arXiv:1211.6453 [hep-ph]].


\bibitem{Huber:2015sra}
  T.~Huber, T.~Hurth and E.~Lunghi,
  ``Inclusive $ \overline{B}\to {X}_s{\ell}^{+}{\ell}^{-} $ : complete angular analysis and a thorough study of collinear photons,''
  JHEP {\bf 1506} (2015) 176
  [arXiv:1503.04849 [hep-ph]].









\bibitem{Aaij:2013qta}
  R.~Aaij {\it et al.} [LHCb Collaboration],
  ``Measurement of Form-Factor-Independent Observables in the Decay $B^{0} \to K^{*0} \mu^+ \mu^-$,''
  Phys.\ Rev.\ Lett.\  {\bf 111} (2013) 191801
  [arXiv:1308.1707 [hep-ex]].



\bibitem{Aaij:2015oid}
  R.~Aaij {\it et al.} [LHCb Collaboration],
  ``Angular analysis of the $B^{0} \to K^{*0} \mu^{+} \mu^{-}$ decay using 3 fb$^{-1}$ of integrated luminosity,''
  JHEP {\bf 1602} (2016) 104
  [arXiv:1512.04442 [hep-ex]].


\bibitem{Hurth:2013ssa}
  T.~Hurth and F.~Mahmoudi,
  ``On the LHCb anomaly in B $\to K^*\ell^+\ell^-$,''
  JHEP {\bf 1404} (2014) 097
  [arXiv:1312.5267 [hep-ph]].

\bibitem{Hurth:2014zja}
  T.~Hurth and F.~Mahmoudi,
  ``Signs for new physics in the recent LHCb data?,''
  Nucl.\ Part.\ Phys.\ Proc.\  {\bf 263-264} (2015) 38
  [arXiv:1411.2786 [hep-ph]].




\bibitem{Ligeti:1997tc}
  Z.~Ligeti, L.~Randall and M.~B.~Wise,
  ``Comment on nonperturbative effects in  $\bar B \to X_s \gamma$,''
  Phys.\ Lett.\ B {\bf 402} (1997) 178
  [hep-ph/9702322].


\bibitem{Chay:1990da}
  Chay J, Georgi H, Grinstein B. 
  ``Lepton energy distributions in heavy meson decays from QCD,''
     Phys.\ Lett.\  B {\bf 247} (1990) 399.
 
 \bibitem{Bigi:1992su}
  Bigi II, Uraltsev NG, Vainshtein AI.
   ``Non-perturbative corrections to inclusive beauty and charm decays: QCD
  versus phenomenological models,''
     Phys.\ Lett.\  B {\bf 293} (1992) 430
   [Erratum-ibid.\  B {\bf 297} (1993) 477]
   [arXiv:hep-ph/9207214].
  
\bibitem{Voloshin:1996gw}
  M.~B.~Voloshin,
  ``Large O (m(c)**-2) nonperturbative correction to the inclusive rate of the decay $\bar B \to X_s \gamma$,''
  Phys.\ Lett.\ B {\bf 397} (1997) 275
  [hep-ph/9612483].




\bibitem{Lee:2006wn}
     S.~J.~Lee, M.~Neubert and G.~Paz,
  ``Enhanced non-local power corrections to the $\bar B \to X_s \gamma$ decay rate,''
     Phys.\ Rev.\  D {\bf 75} (2007) 114005
   [arXiv:hep-ph/0609224].



\bibitem{Benzke:2010js}
  M.~Benzke, S.~J.~Lee, M.~Neubert and G.~Paz,
  ``Factorization at Subleading Power and Irreducible Uncertainties in $\bar B\to X_s\gamma$ Decay,''
  JHEP {\bf 1008} (2010) 099
  [arXiv:1003.5012 [hep-ph]].




\bibitem{Beneke:2002ph}
  M.~Beneke, A.~P.~Chapovsky, M.~Diehl and T.~Feldmann,
  ``Soft collinear effective theory and heavy to light currents beyond leading power,''
  Nucl.\ Phys.\ B {\bf 643} (2002) 431
  [hep-ph/0206152].








\bibitem{Lee:2005pk}
     K.~S.~M.~Lee and I.~W.~Stewart,
  ``Shape-function effects and split matching in $\bar B \to  X_s \ell^+\ell^-$,''
     Phys.\ Rev.\ D {\bf 74} (2006) 014005
   [hep-ph/0511334].

\bibitem{Lee:2005pw}
     K.~S.~M.~Lee, Z.~Ligeti, I.~W.~Stewart and F.~J.~Tackmann,
  ``Universality and $M_X$  cut effects in $\bar B \to  X_s \ell^+\ell^-$,''
     Phys.\ Rev.\ D {\bf 74} (2006) 011501
   [hep-ph/0512191].


\bibitem{Lee:2008xc}
      K.~S.~M.~Lee and F.~J.~Tackmann,
  ``Non-perturbative $m_X$ cut effects in $\bar B \to  X_s \ell^+\ell^-$ observables,''
     Phys.\ Rev.\  D {\bf 79} (2009) 114021
   [arXiv:0812.0001 [hep-ph]].



\bibitem{Buchalla:1995vs}
  G.~Buchalla, A.~J.~Buras and M.~E.~Lautenbacher,
  ``Weak decays beyond leading logarithms,''
  Rev.\ Mod.\ Phys.\  {\bf 68} (1996) 1125
  [hep-ph/9512380].


 \bibitem{Beneke:2009az}
    M.~Beneke, G.~Buchalla, M.~Neubert and C.~T.~Sachrajda,
  ``Penguins with Charm and Quark-Hadron Duality,''
   Eur.\ Phys.\ J.\  C {\bf 61} (2009) 439
   [arXiv:0902.4446 [hep-ph]].




\bibitem{Bauer:2001yt}
  C.~W.~Bauer, D.~Pirjol and I.~W.~Stewart,
  Phys.\ Rev.\ D {\bf 65} (2002) 054022
  [hep-ph/0109045].



\bibitem{Benzke:2017woq}
  M.~Benzke, T.~Hurth and S.~Turczyk,
  ``Subleading power factorization in $ \bar{B}\to {X}_s{\ell}^{+}{\ell}^{-} $,''
  JHEP {\bf 1710} (2017) 031
  [arXiv:1705.10366 [hep-ph]].







\end{thebibliography}
\end{document}